\documentclass[journal]{IEEEtran}
\usepackage{stmaryrd}
\usepackage{amsfonts}
\usepackage{amsmath}
\usepackage{amssymb}
\usepackage{cite}
\usepackage{graphicx}
\usepackage{subfigure}
\usepackage{amsthm}
\usepackage{algorithmic}
\usepackage{algorithm}
\usepackage{stfloats}
\usepackage{textcomp}

\newtheorem{Theorem}{Theorem}

\newtheorem{Example}{Example}

\def\QED{\IEEEQED\vspace{0.1in}}
\newcommand{\nchoosek}[2]{\binom{#1}{#2}}

\def\x{\mathbf{x}}

\ifCLASSINFOpdf
\else
\fi
\begin{document}
%
\title{Balanced Modulation for Nonvolatile Memories}

%

\author{Hongchao~Zhou,
        Anxiao~(Andrew)~Jiang,~\IEEEmembership{Member,~IEEE,}
        Jehoshua~Bruck,~\IEEEmembership{Fellow,~IEEE}
\thanks{This work was supported in part by the NSF CAREER Award CCF-0747415, the NSF
grant ECCS-0802107, and by an NSF-NRI award. This paper was presented in part at IEEE International Symposium on Information Theory (ISIT), St. Petersburg, Russia, August 2011.}
\thanks{H. Zhou and J. Bruck are with the Department
of Electrical Engineering, California Institute of Technology, Pasadena, CA, 91125.
{\it Email: hzhou@caltech.edu, bruck@caltech.edu}}
\thanks{A. Jiang is with the Computer Science and Engineering Department,
Texas A\&M University,
College Station, TX 77843. {\it Email: ajiang@cse.tamu.edu}}
}
\maketitle
%
%

%

\begin{abstract}This paper presents a practical writing/reading scheme in nonvolatile memories, called balanced modulation, for minimizing the asymmetric component of errors. The main idea is to encode data using a balanced error-correcting code. When reading information from a block, it adjusts the reading threshold such that the resulting word is also balanced or approximately balanced. Balanced modulation has suboptimal performance for any cell-level distribution and it can be easily implemented in the current systems of nonvolatile memories.
Furthermore, we studied the construction of balanced error-correcting codes, in particular, balanced LDPC codes.
It has very efficient encoding and decoding algorithms, and it is more efficient than prior construction of balanced error-correcting codes.
\end{abstract}





%
\IEEEpeerreviewmaketitle

\section{Introduction}
%
%
%
%
\IEEEPARstart{N}{onvolatile} memories, like EPROM, EEPROM, Flash memory
or Phase-change memory (PCM), are memories that can keep the data content
even without power supply. This property enables them to be
used in a wide range of applications, including cellphones,
consumers, automotive and computers. Many research
studies have been carried out on nonvolatile memories because of their unique features, attractive applications and huge marketing
demands.

An important challenge for most nonvolatile memories
is data reliability. The stored data can be lost due to many
mechanisms, including cell heterogeneity, programming noise, write disturbance, read disturbance, etc. \cite{Bez03,Parovano04}.
From a long-term view, the change in data has an asymmetric property. For example,
the stored data in flash memories is represented by the voltage levels of transistors,
which drift in one direction because of charge leakage. In PCM, another class of nonvolatile memories,
the stored data is determined by the electrical resistance of the cells, which drifts due to
thermally activated crystallization of the amorphous
material \cite{Wong2010}. All these mechanisms make the errors in nonvolatile memories be heterogeneous, asymmetric, time dependent and unpredictable. These properties bring substantial difficulties to researchers attempting to develop simple and efficient error-correcting schemes.

\begin{figure}[!t]
\centering
\includegraphics[width=3.6in]{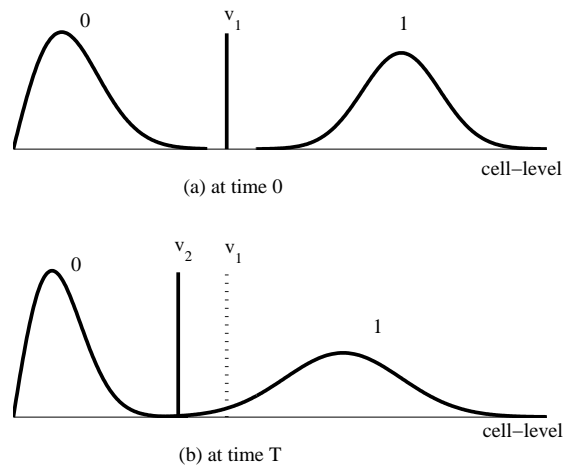}
\caption{An illustration of the voltage distributions for bit ``1" and bit ``0" in flash memories.}
\label{fig_voltageDistribution}
\end{figure}

To date, existing coding schemes for nonvolatile memories commonly use fixed thresholds to read data.
For instance, in flash memories, a threshold voltage level $\mathbf{v}$ is predetermined; when reading data from a cell,
it gets `1' if the voltage level is higher than $\mathbf{v}$, and otherwise it gets `0'. To
increase data reliability, error-correcting codes such as
Hamming code, BCH code, Reed-Solomon code and LDPC
code are applied in nonvolatile memories to combat errors.
Because of the asymmetric feature of nonvolatile memories, a fixed threshold usually introduces too many asymmetric errors after a long duration \cite{Mielke2008}, namely,
the number of $1\rightarrow 0$ errors is usually much larger than the number of $0\rightarrow 1$ errors.
To overcome the limitations of fixed thresholds in reading data in nonvolatile memories, dynamic thresholds
are introduced in this paper. To better understand this, we use flash memories for illustration, see Fig. \ref{fig_voltageDistribution}.
The top figure is for newly written data, and the bottom figure is for old data that has been stored for a long time $T$.
In the figures, assume the left curve
indicates the voltage distribution for bit `0' (a bit `0' is written during programming) and
the right curve indicates the voltage distribution for bit `1'. At time $0$ (the moment after programming),
it is best to set the threshold voltage as $\mathbf{v}=v_1$, for separating bit `1' and `0'.
But after a period of time, the voltage distribution will change. In this case, $v_1$ is no longer the best choice, since it will
introduce too many $1\rightarrow 0$ errors. Instead, we can set the threshold voltage
as $\mathbf{v}=v_2$ (see the second plot in the figure), to minimize the error probability. This also applies to other nonvolatile memories, such as PCMs.

Although best dynamic reading thresholds lead to much less errors than fixed ones, certain difficulties exist in
determining their values at a time $t$. One reason is  that the accurate level distributions for bit `1' and `0' at any the current time
are hard to obtain due to the lack of time records,
the heterogeneity of blocks, and the unpredictability of exceptions.  Another possible method is to classify
all the cell levels into two groups based on unsupervised clustering and then map them into `1's and `0's. But when the border between bit `1's and `0's becomes fuzzy, mistakes of clustering may cause significant number of
reading errors.
In view of these considerations, in this paper,
we introduce a simple and practical writing/reading scheme in nonvolatile memories, called \emph{balanced modulation}, which
is based on the construction of balanced codes (or balanced error-correcting codes) and it aims to minimize the asymmetric component of errors
in the current block.

Balanced codes, whose codewords have an equal number of $1$s and $0$s, have been studied in several literatures.
Knuth, in 1986, proposed a simple method of constructing balanced codes \cite{Knuth86}.
 In his method, given an information
word of $k$-bits ($k$ is even), the encoder inverts the first $i$ bits such
that the modified word has an equal number of $1$s and $0$s.
Knuth showed that such an integer $i$ always exists, and it is represented
by a balanced word of length $p$. Then a codeword consists of
an $p$-bit prefix word and an $k$-bit modified information word.
For decoding, the decoder can easily retrieve the value of $i$ and then
get the original information word by inverting the first $i$ bits of the $k$-bit information word again. Knuth's method was later
improved or modified by many researchers \cite{Al-Bassam90,Tallini96,Web2010,Immink2010}.
Based on balanced codes, we have a scheme of balanced modulation.  It encodes the stored data as balanced codewords; when reading data from a block,
it adjusts the reading threshold dynamically such that the resulting word to read is also balanced (namely, the number of
1s is equal to the number of 0s) or approximately balanced. Here, we call this dynamic reading threshold as a \emph{balancing threshold}.

There are several benefits of applying balanced modulation in nonvolatile memories. First, it increases
the \emph{safety gap} of  $1$s and $0$s. With a fixed threshold,  the \emph{safety gap}
is determined by the minimum difference between cell levels and the threshold. With balanced modulation, the safety gap is
the minimum difference between cell levels for $1$ and those for $0$. Since the cell level for an individual
cell has a random distribution due to the cell-programming
noise \cite{Brewer2008,Lue2008}, the actual value of the charge level varies
from one write to another. In this case, balanced modulation is more robust than the commonly used fixed-threshold approach in combating
programming noise. Second, as we discussed, balanced modulation can is a very simple solution that minimizes the influence of cell-level drift.
It was shown in \cite{Cai2012} that cell-level drift in flash memories introduces the most dominating errors. Third, balanced
modulation can efficiently reduce errors introduced by some other mechanisms, such as the change of external temperatures and the current leakage of
other reading lines, which result in the shift of cell levels in a same direction.
Generally,
balanced modulation is a simple approach that minimizes the influence of noise
asymmetries, and it can be easily implemented on current memory devices without hardware changes.
The balanced condition on codewords enables us to select a much better threshold dynamically than the commonly used fixed threshold
when reading data from a block.

The main contributions of the paper are
\begin{enumerate}
\item We study balanced modulation as a simple, practical and efficient approach to minimize asymmetric component of errors in nonvolatile memories.
\item A new construction of balanced error-correcting codes, called balanced LDPC code, is introduced and analyzed, which has a higher rate than prior constructions.
\item We investigate partial-balanced modulation, for its simplicity of constructing error-correcting codes, and then we extend our discussions from binary cells to multi-level cells.
\end{enumerate}

\section{Scope of This Paper}

\subsection{Performance and Implementation}

In the first part of this paper, including Section \ref{section_bal_balancedscheme}, Section \ref{section_bal_biterrorrate} and Section \ref{section_bal_implementation}, we focus on the introduction and performance of balanced modulation. In particular, we demonstrate that
balanced modulation introduces much less errors than the traditional approach based on fixed thresholds.
For any cell-level distributions, the balancing threshold used in balanced modulation is suboptimal among all the possible reading thresholds, in the term of total number of errors.  It enables balanced modulation to be
adaptive to a variety of channels characters, hence, it makes balanced modulation applicable for most types of nonvolatile memories.
Beyond storage systems, balanced modulation can also be used in optimal communication, where the strength of received signals shifts due to many factors like the transmitting
distance, temperature, etc.

A practical and very attractive aspect of balanced modulation is that it can be easily implemented in the current systems of nonvolatile memories. The only change is that, instead of
using a fixed threshold in reading a binary vector, it allows this threshold to be adaptive. Fortunately, this operation can be implemented physically, making the process of data reading reasonably fast.
In this case, the reading process is based on hard decision.

If we care less about reading speed, we can have soft-decision decoding, namely, reading data without using a threshold.
We demonstrate that the prior knowledge that the stored codeword is balanced is very useful. It helps us to better estimate the
current cell-level distributions, hence, resulting in a better performance in bit error rate.

\subsection{Balanced LDPC Code}

Balanced modulation can efficiently reduce bit error rate when reading data from a block.
A further question is how to construct balanced codes that are capable of correcting errors. We call such codes \emph{balanced error-correcting codes}.
Knuth's method cannot correct errors. In \cite{vanTilborg89}, van Tilborg and Blaum presented a family
of balanced binary error-correcting codes. The idea is to consider balanced blocks as symbols over an alphabet and to
construct error-correcting codes over that alphabet by concatenating $n$ blocks of length $2l$ each. Due to the constraint in the code construction,
this method achieves only moderate rates. Error-correcting balanced codes with higher rates were presented by Al-Bassam and Bose in \cite{Al-Bassam90},
however, their construction considers only the case that the number of errors is at most $4$. In \cite{Mazumdar2009}, Mazumdar, Roth, and Vontobel studied
linear balancing sets, namely, balancing sets that are linear subspaces $\mathbb{F}^n$, which are applied in obtaining coding schemes that combine balancing
and error correction. Recently, Weber,  Immink and Ferreira extent Knuth's method to let it equipped with error-correcting capabilities \cite{Weber2012}. Their idea is to assign different error protection levels to the prefix and modified information word in Knuth's construction.  So their construction is a concatenation of two error-correct codes with different error correcting capabilities.
In Section \ref{section_bal_LDPC}, we introduce a new construction of balanced error-correcting codes, which is based on LDPC code, so called balanced LDPC code. Such a construction
has a simple encoding algorithm and its decoding complexity based on message-passing algorithm is asymptotically equal to the decoding complexity of the original (unbalanced) LDPC code. We demonstrate that balanced LDPC code
has error-correcting capability very close to the original (unbalanced) LDPC code.

\subsection{Partial-Balanced Modulation and Its Extension}

Our observation is that the task of constructing efficient balanced error-correcting codes with simple encoding and decoding algorithms is
not simple, but it is much easier to construct error-correcting codes that are partially balanced, namely,
only a certain segment (or subsequence) of
 each codeword is balanced. Motivated by this observation, we propose a variant of balanced modulation, called partial-balanced modulation. When
  reading from a block, it adjusts the reading threshold such that
the segment of the resulting word is balanced. Partial-balanced modulation has a performance very close to that of balanced modulation,
and it has much simpler constructions of error-correcting codes than balanced modulation. Another question that we address in the third part
is how to extend the scheme of balanced modulation or partial-balanced modulation to be used in nonvolatile memories with multi-level cells.
Details will be provided in Section \ref{section_bal_variant} and Section \ref{section_bal_multicell}.

\section{Balanced Modulation}
\label{section_bal_balancedscheme}

\begin{figure}[!t]
\centering
\includegraphics[width=3.6in]{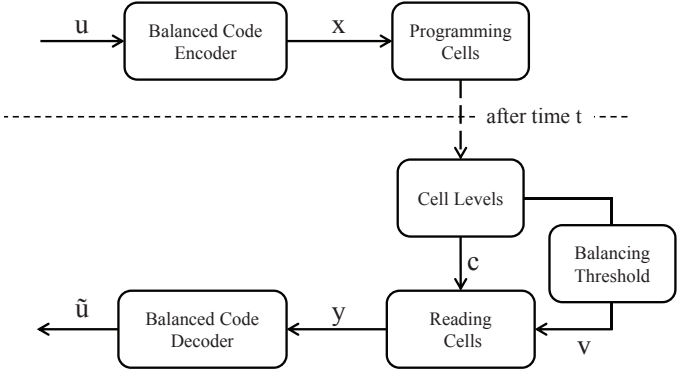}
\caption{The diagram of balanced modulation.}
\label{fig_balancedmodulation}
\end{figure}

For convenience, we consider different types of nonvolatile memories in the same framework where data is represented by cell levels, such as voltages in flash memories and resistance in phase-change memories.
The scheme of balanced modulation is sketched in Fig.~\ref{fig_balancedmodulation}. It can be divided into two steps: programming step and
reading step.

(1) In the programming step, we encode data based a balanced (error-correcting) code. Let $k$ denote the dimension of
the code and $n$ denote the number of cells in a block, then given  a message $\mathbf{u}\in \{0,1\}^n$,
it is mapped to a balanced codeword $\mathbf{x}\in \{0,1\}^n$ such that $|\mathbf{x}|=\frac{n}{2}$ where $|\mathbf{x}|$ is the Hamming weight of $\mathbf{x}$.

(2) In the reading step, we let $\mathbf{c}=c_1c_2...c_n \in \mathcal{R}^n$ be the current levels of the $n$ cells to read. A balancing threshold $\mathbf{v}$ is determined
based on $\mathbf{c}$ such that the resulting word, denoted by $\mathbf{y}=y_1y_2...y_n$, is also balanced, namely, $|\mathbf{y}|=\frac{n}{2}$. For each $i\in \{1,2,...,n\}$,
$y_i=1$ if and only if $c_i\geq \mathbf{v}$, otherwise $y_i=0$.  By applying the decoder of the balanced (error-correcting) code, we
get a binary output $\mathbf{\tilde{u}}$, which is the message that we read from the block.

\begin{figure}[!ht]
\centering
\includegraphics[width=3.6in]{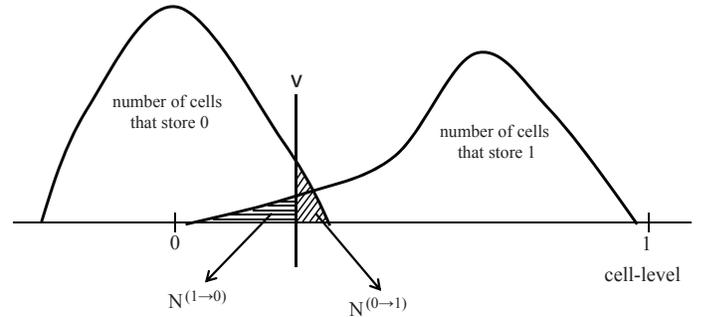}
\caption{Cell-level distributions for $1$ and $0$, and the reading threshold.}
\label{fig_balancingthreshold}
\end{figure}

Let us intuitively understanding the function of balanced modulation based on the demonstration of Fig.~\ref{fig_balancingthreshold},
which depicts the cell-level distributions for those cells that store $0$ or $1$. Given a reading threshold $\mathbf{v}$, we use $N^{(1\rightarrow0)}$
denote the number of $1\rightarrow 0$ errors and use $N^{(0\rightarrow 1)}$ denote the number of $0\rightarrow 1$ errors, as the tails
marked in the figure. Then
$$N^{(1\rightarrow0)}=|\{i:x_i=1, y_i=0\}|,$$
$$N^{(0\rightarrow1)}=|\{i:x_i=0, y_i=1\}|.$$

We are ready to see
$$|\mathbf{y}|=|\mathbf{x}|-N^{(1\rightarrow0)}+N^{(0\rightarrow1)},$$
where $|\mathbf{x}|$ is the Hamming weight of $\mathbf{x}$.

According to the definition, a balancing threshold is the one that makes $\mathbf{y}$ being balanced, hence,
$$N^{(1\rightarrow0)}(\mathbf{v})=N^{(0\rightarrow1)}(\mathbf{v}),$$
i.e., a balancing threshold results in the same number of $1\rightarrow 0$ errors and $0\rightarrow 1$ errors.

We define $N_e(\mathbf{v})$ as the total number of errors based on a reading threshold $\mathbf{v}$, then
$$N_e(\mathbf{v})=N^{(1\rightarrow 0)}(\mathbf{v})+ N^{(0\rightarrow 1)}(\mathbf{v}).$$
If the cell-level distributions for those cells that store $1$ and those cells that store $0$ are known, then the balancing threshold may not be the best reading threshold that we can have, i.e.,
$N_e(\mathbf{v})$ may not be minimized based on the balancing threshold. Let $v_b$ denote the balancing threshold, as a comparison,
we can have an optimal threshold $v_o$, which is defined by
$$v_o=\arg\min_{\mathbf{v}}N_e(\mathbf{v}).$$
Unfortunately, it is almost impossible for us to know the cell-level distributions for those cells that store $1$ and those cells that store $0$ without knowing
the original word $\mathbf{x}$. From this sense, the optimal threshold $v_o$ is imaginary. Although we are not able to determine $v_o$,
the following result shows that the balancing threshold $v_b$ has performance comparable to that of $v_o$. Even in the worst case, the number
of errors introduced based on $v_b$ is at most two times that
introduced by $v_o$, implying the suboptimality of the balancing threshold $v_b$.

\begin{Theorem} Given any balanced codeword $\mathbf{x}\in \{0,1\}^n$ and cell-level vector $\mathbf{c}\in \mathcal{R}^n$, we have
$$N_e(v_b)\leq 2N_e(v_o).$$
\end{Theorem}

\proof Given the balancing threshold $v_b$, the number of $0\rightarrow 1$ errors equals the number of $1\rightarrow 0$ errors, hence, the total number
of errors is
$$N_e(v_b)= 2N^{(1\rightarrow 0)}(v_b)=2N^{(0\rightarrow 1)}(v_b).$$

If $v_o\geq v_b$, the number of $1\rightarrow 0$ errors
$N^{(1\rightarrow 0)}(v_o)\geq N^{(1\rightarrow 0)}(v_b)$. Therefore, $$N_e(v_b)\leq 2N^{(1\rightarrow 0)}(v_o)\leq 2N_e(v_o).$$

Similarly, if $v_o<v_b$, by considering only $0\rightarrow 1$ errors, we get the same conclusion.
\hfill\QED

\begin{figure}[!ht]
\centering
\includegraphics[width=3.6in]{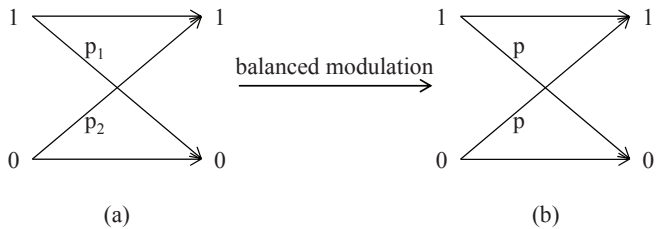}
\caption{Balanced modulation to turn a binary asymmetric channel with crossover probabilities $p_1>p_2$ into a binary symmetric channel with $p_2<p<p_1$.}
\label{fig_channelchange}
\end{figure}

Now we compare the balancing threshold $v_b$ with a fixed threshold, denoted by $v_f$. As shown in Fig.~\ref{fig_balancingthreshold},
if we set the reading threshold as fixed $v_f=\frac{1}{2}$, then it will
introduce much more errors then the balancing threshold. Given a fixed threshold $v_f$, after a long duration, we can characterize
the storage channel as a binary asymmetric channel, as shown in Fig.~\ref{fig_channelchange}(a), where $p_1>p_2$.
Balanced modulation is actually a process of modifying the channel to make it being symmetric. As
a result, balanced modulation results in a binary symmetric channel with crossover probability $p$ such that $p_2<p<p_1$. When $p_2\ll p_1$,
it has $p-p_2\ll p_1-p$. In this case, the bit error rate is reduced from
$\frac{p_1+p_2}{2}$ to $p$, where $p\ll \frac{p_1+p_2}{2}$.

\section{Bit-Error-Rate Analysis}
\label{section_bal_biterrorrate}

To better understand different types of reading thresholds as well as their performances, we study them from the expectation (statistical) perspective.
Assume that we write $n$ bits (including $k$ ones) into a block at time $0$, let $g_t(v)$ denote the probability density function  (p.d.f.) of the cell level at time $t$ that stores a bit $0$, and let $h_t(v)$ denote the p.d.f. of the cell level at time $t$
that stores $1$. Then at time $t$, the bit error rate of the block based on a reading threshold $\mathbf{v}$ is given by
$$p_e(\mathbf{v})=\frac{1}{2}\int_{\mathbf{v}}^{\infty} g_t(u) du+ \frac{1}{2}\int_{-\infty}^{\mathbf{v}} h_t(v) dv.$$

According to our definition, a  balancing threshold $v_b$ is chosen such that $N^{(1\rightarrow0)}(v_b)=N^{(0\rightarrow\infty)}(v_b)$, i.e., the number of $1\rightarrow 0$ errors is equal to the number of $0\rightarrow 1$ errors. As the block length $n$ becomes sufficiently large,
we can approximate $N^{(1\rightarrow0)}(v_b)$ as $\frac{n}{2}\int_{-\infty}^{\mathbf{v}} h_t(v) dv$ and
approximate $N^{(0\rightarrow\infty)}(v_b)$ as $\frac{n}{2}\int_{\mathbf{v}}^{\infty} g_t(u) du$.
So when $n$ is large, we approximately have
$$\int_{v_b}^{\infty} g_t(u) du = \int_{-\infty}^{v_b} h_t(v) dv.$$

Differently, an optimal reading threshold $v_o$ is the one that minimizes the total number of errors.
When $n$ is large, we approximately have
$$v_o=\arg\min_{\mathbf{v}} p_e(\mathbf{v}).$$
When $g_t(v)$ and $h_t(v)$ are continuous functions, the solutions of $v_o$ are
$$v_o=\pm \infty\textrm{ or } g_t(v_o)= h_t(v_o).$$
That means $v_o$ is one of the intersections of $g_t(v)$ and $h_t(v)$ or one of the infinity points.

Generally, $g_t(v)$ and $h_t(v)$ are various for different nonvolatile memories and different blocks, and they have different dynamics over time.
It is not easy to find a perfect model to characterize $g_t(v)$ and $h_t(v)$, but there are two trends about them in timescale. The change of a cell level can be treated as a superposition of these two trends. First, due to cell-level drift, the difference between the means of $g_t(v)$ and $h_t(v)$ becomes smaller. Second, due to the existence of different types of noise and
disturbance, their variances increases over time. To study the performance of balanced modulation, we consider both of the effects separately in some simple scenarios.

\begin{figure}[!ht]
\centering
\includegraphics[width=3.6in]{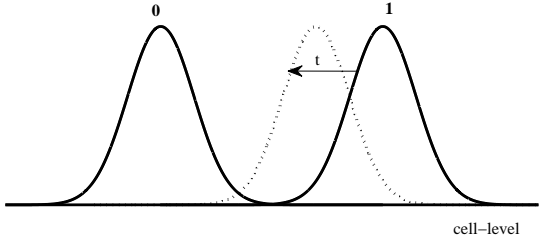}
\caption{An illustration of the first model
 with $g_t(v)=\mathcal{N}(0, \sigma)$ and $h_t(v)=\mathcal{N}(1-t, \sigma)$.}
\label{fig_balancedexample1}
\end{figure}

\begin{Example}
Let $g_t(v)=\mathcal{N}(0, \sigma)$ and $h_t(v)=\mathcal{N}(1-t, \sigma)$, as illustrated in Fig.~\ref{fig_balancedexample1}. We assume that the fixed threshold is $v_f=\frac{1}{2}$, which
satisfies $g_0(v_f)=h_0(v_f)$.
\end{Example}

In the above example, the cell-level distribution corresponding to bit `1' drifts but its variance does not change. We have
$$v_b=v_o=\frac{1-t}{2},\quad v_f=\frac{1}{2}.$$

At time $t$, the bit error rate based on a reading threshold $\mathbf{v}$
is
$$p_e(\mathbf{v})=\frac{1}{2}\Phi(-\frac{\mathbf{v}}{\sigma})+ \frac{1}{2}\Phi(-\frac{1-t-\mathbf{v}}{\sigma}),$$
where
$\Phi(x)=\frac{1}{\sqrt{2\pi}} \int_{-\infty}^{x}e^{-t^2/2}dt$.

\begin{figure}[!ht]
\centering
\includegraphics[width=3.6in]{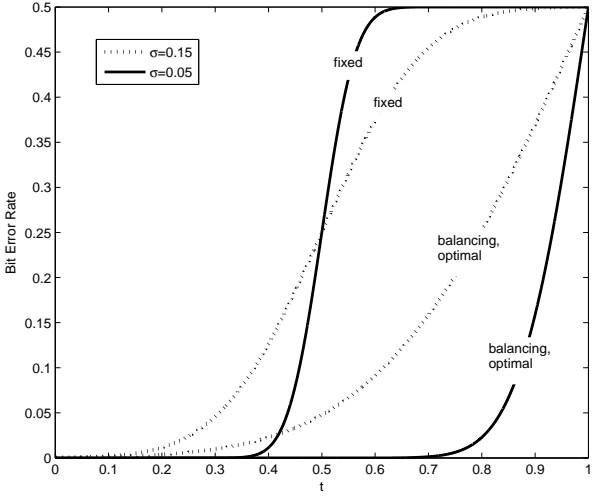}
\caption{Bit error rates as functions of time $t$, under the first model
 with $g_t(v)=\mathcal{N}(0, \sigma)$ and $h_t(v)=\mathcal{N}(1-t, \sigma)$.}
\label{fig_errorProbability2}
\end{figure}

For different selections of reading thresholds, $p_e(\mathbf{v})$ is plotted in Fig.~\ref{fig_errorProbability2}. It shows
that the balancing threshold and the optimal threshold have the same performance, which is much better than the performance of a fixed threshold.
When cell levels drift, balanced modulation can significantly reduce the bit error rate of a block.

\begin{figure}[!ht]
\centering
\includegraphics[width=3.6in]{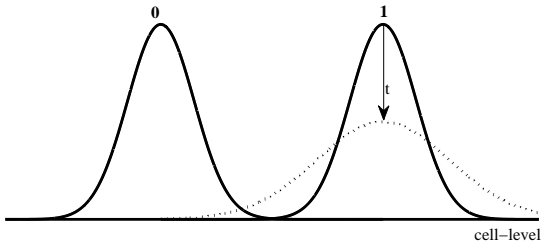}
\caption{An illustration of the second model
 with $g_t(v)=\mathcal{N}(0, \sigma)$ and $h_t(v)=\mathcal{N}(1, \sigma+t)$.}
\label{fig_balancedexample2}
\end{figure}

\begin{Example}
Let $g_t(v)=\mathcal{N}(0, \sigma)$ and $h_t(v)=\mathcal{N}(1, \sigma+t)$, as illustrated in Fig.~\ref{fig_balancedexample2}. We assume that the fixed threshold is $v_f=\frac{1}{2}$, which
satisfies $g_0(v_f)=h_0(v_f)$.
\end{Example}

In this example, the variance of the cell-level distribution corresponding to bit `1' increases as the time $t$ increases. We have
$$e^{-\frac{{v_o}^2}{2\sigma^2}}=\frac{\sigma}{\sigma+t}e^{-\frac{(1-v_o)^2}{2(\sigma+t)^2}}, \quad v_b=\frac{1}{2+t/\sigma}, \quad v_f=\frac{1}{2}.$$

\begin{figure}[!ht]
\centering
\includegraphics[width=3.6in]{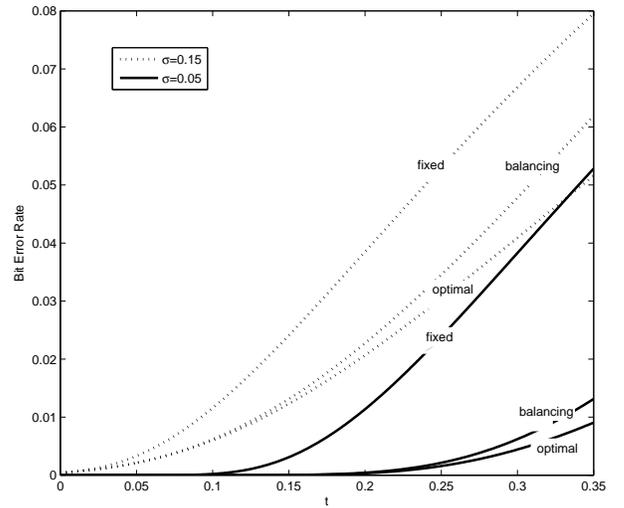}
\caption{Bit error rates as functions of time $t$, under the second model
 with $g_t(v)=\mathcal{N}(0, \sigma)$ and $h_t(v)=\mathcal{N}(1, \sigma+t)$.}
\label{fig_errorProbability1}
\end{figure}

At time $t$,  the bit error rate based on a threshold $\mathbf{v}$
is
$$p_e(\mathbf{v})=\frac{1}{2}\Phi(-\frac{\mathbf{v}}{\sigma})+ \frac{1}{2}\Phi(-\frac{1-\mathbf{v}}{\sigma+t}),$$
which is plotted in Fig.~\ref{fig_errorProbability1} for different thresholds. It shows that balancing thresholds introduce much less errors
than fixed thresholds when bit `1' and `0' have different reliability (reflected by their variances), although they introduce slightly
more errors than optimal thresholds.

In practice, the cell-level distributions at a time $t$ are much more complex
than the simple Gaussian distributions, and the errors introduced are due to many complex mechanisms. However, the above analysis based two simple models are still useful,  because they reflect the trends of the cell level changes, which
is helpful for analyzing the time-dependent errors in nonvolatile memories.

\section{Implementation}
\label{section_bal_implementation}

Balanced modulation can be easily implemented on the current architecture of nonvolatile memories.
The process described in the previous sections can be treated as a hard decision approach, where
a reading threshold is selected to separate all the cell levels as zeros and ones. In this section,
we discuss a few methods of determining balancing thresholds quickly, as well as their implementations in nonvolatile memories.
Furthermore,  we discuss soft decision implementation of balanced modulation, namely, we do not read data based on
a reading threshold, and the decoder can get access into all the cell levels (cell-level vector $\mathbf{c}$) directly. In this case,
we want to know how the prior information that the stored codeword is balanced can help us to increase the success rate of decoding.

\subsection{Balancing Threshold for Hard Decision}

Given a block of $n$ cells, assume their current levels are $\mathbf{c}=c_1c_2...c_n$. Our problem
is to determine a threshold $v_b$ such that there are $\frac{n}{2}$ cells or approximately $\frac{n}{2}$ cells will be read as ones.
A trivial method is to sort all the $n$ cell levels in the decreasing order such that
$c_{i_1}\geq c_{i_2} \geq ...\geq c_{i_n}$. Then $v_b=\frac{c_{i_k}+c_{i_{k+1}}}{2}$ is our desired balancing
threshold. The disadvantage of this method is that it needs $O(n\log n)$ computational time, which may slow
down the reading speed when $n$ is large. To reduce the reading time, we hope that the balancing threshold can be controlled by hardware.

Half-interval search is a simple approach of determining the balancing threshold. Assume
it is known that $v_b$ is $\in [l_1, l_2]$ with $l_1<l_2$. First,
we set the reading threshold as $\frac{l_1+l_2}{2}$, based on which a simple circuit can quickly
detect the number of ones in the resulting word, denoted by $k$.
If $k<\frac{n}{2}$, we reset the interval $[l_1,l_2]$
as $[l_1, \frac{l_1+l_2}{2}]$. If $k>\frac{n}{2}$, we reset the interval $[l_1,l_2]$ as
$[\frac{l_1+l_2}{2}, l_2]$. Then we repeat this procedure until we get a reading threshold such that
$k=\frac{n}{2}$ or $l_2-l_1\leq \epsilon$ for a reading precision $\epsilon$.

\subsection{Relaxed Balancing Threshold}

Half-interval search is an iterative approach of determining the balancing threshold such that
the resulting word is well balanced. To further reduce the reading time, we can relax the constraint
on the weight of the resulting word, namely, we can let the number of ones in the resulting word be approximately
$\frac{n}{2}$, instead of accurately $\frac{n}{2}$.

For instance, we can simply set the balancing threshold as
$$v_b=\frac{\sum_{i=1}^n c_i}{n}=\textrm{mean}(\mathbf{c}).$$
Obviously, such $v_b$ reflects the cell-level drift and it can be easily implemented by a simple circuit.

More precisely, we can treat $\textrm{mean}(\mathbf{c})$ as the first-order approximation,
in this way, we write $v_b$ as
$$v_b=\textrm{mean}(\mathbf{c})+a(\frac{1}{2}-\textrm{mean}(\mathbf{c}))^2,$$
where $a$ is a constant depending on the noise model of memory devices.

\subsection{Prior Probability for Soft Decision}

\label{subsection_soft}

Reading data based on hard decision is preferred in nonvolatile memories, regarding to its advantages in reading speed and computational complexity
compared to soft decision decoding.
However, in some occasions, soft decision decoding is still useful for increasing the decoding success rate.
We demonstrate that the prior knowledge that the stored codewords are balanced can help us to better estimate
the cell-level probability distributions for $0$ or $1$.  Hence, it leads to a better soft decoding performance.

We assume that given a stored bit, either $0$ or $1$, its cell level is Gaussian distributed. (We may also use some other distribution models according to the physical properties of memory devices, and our goal is to have a better estimation of model parameters). Specifically, we assume that the cell-level probability distribution for $0$ is
$\mathcal{N}(u_0, \sigma_0)$ and the cell-level probability distribution for $1$ is
$\mathcal{N}(u_1, \sigma_1)$. Since the codewords are balanced, the probability for a cell being $0$ or $1$ is equal.
So we can describe cell levels by a Gaussian Mixture Model. Our goal is to find the maximum likelihood $u_0, \sigma_0, u_1, \gamma_1$
based on the cell-level vector $\mathbf{c}$, namely, the parameters that maximize
$$P(\mathbf{c}|u_0, \sigma_0, u_1, \sigma_1).$$

Expectation-Maximization (EM) algorithm is an iterative method that can easily find the maximum likelihood $u_0, \sigma_0, u_1, \gamma_1$.
The EM iteration alternates between performing an expectation (E) step and a maximization (M) step.
Let $\mathbf{x}=x_1x_2...x_n$ be the codeword stored in the current block, and let $\lambda_t=[u_0(t), \sigma_0(t), u_1(t), \gamma_1(t)]$ be the estimation
of the parameters in the $t$th iteration. In the E-step, it computes the probability for each cell being $0$ or $1$ based on the current estimation
of the parameters, namely, for all $i\in \{1,2,...,n\}$, it computes
$$P(x_i=k|c_i,\lambda_t)=\frac{\frac{1}{\sigma_k(t)}e^{-\frac{(c_i-u_k(t))^2}{2\sigma_k(t)^2}}}{\sum_{k=0}^1\frac{1}{\sigma_k(t)}e^{-\frac{(c_i-u_k(t))^2}{2\sigma_k(t)^2}}}.$$
In the M-step, it computes parameters maximizing the likelihood with given the probabilities obtained in the E-step. Specifically, for $k\in\{0,1\}$,
$$u_k(t+1)=\frac{\sum_{i=1}^n P(x_i=k|c_i,\lambda_t)c_i}{\sum_{i=1}^n P(x_i=k|c_i,\lambda_t)},$$
$$\sigma_k(t+1)^2=\frac{\sum_{i=1}^n P(x_i=k|c_i,\lambda_t)(c_i-u_k(t+1))^2}{\sum_{i=1}^n P(x_i=k|c_i,\lambda_t)}.$$
These estimations of parameters are then used to determine the distribution of $x_i$ in the next E-step.

Assume $u_0, \sigma_0, u_1, \sigma_1$ are the maximum-likelihood parameters, based on which we can calculate the log-likelihood for each variable $x_i$, that
is
$$\lambda_i= \frac{\log f(c_i|x_i=0)}{\log f(c_i|x_i=1)}=\frac{\log\frac{1}{\sigma_0}-\frac{(c_i-u_0)^2}{2\sigma_0^2}}{\log\frac{1}{\sigma_1}-\frac{(c_i-u_1)^2}{2\sigma_1^2}},$$
where $f$ is the probability density function. Based on the log-likelihood of each variable $x_i$, some soft decoding algorithms can be applied to read data, including message-passing algorithms \cite{McEliece98}, linear programming \cite{Feldman2005}, etc. It will be further discussed in the next section for decoding balanced LDPC code.

\section{Balanced LDPC Code}
\label{section_bal_LDPC}

Balanced modulation can significantly reduce the bit error rate of a block in nonvolatile memories, but error correction is still necessary. So we study the construction of balanced error-correcting codes. In the programming step, we encode the information based on a balanced error-correcting code and write it into a block. In the reading step, the reading threshold is adjusted such that it yields a balanced word, but probably erroneous. Then we pass this word to the decoder to further retrieve the original information.

\subsection{Construction}

In this section, we introduce a simple construction of balanced error-correcting  codes, which is based on LDPC codes, called \emph{balanced LDPC code}. LDPC codes, first introduced by
Gallager \cite{Gallager62} in 1962 and rediscovered in 1990s, achieve near
Shannon-bound performances and allow reasonable decoding
complexities. Our construction of balanced LDPC code is obtained by inverting the first $i$ bits of each codeword in a LDPC code
such that the codeword is  balanced, where $i$ is different for different codewords. It is based on Knuth's observation \cite{Knuth86}, that is, given an arbitrary binary word of length $k$ with $k$ even, one can always find an integer $i$ with $0\leq i<k$ such that by inverting the first $i$ bits the word becomes balanced.
Different from the current construction in \cite{Weber2012}, where $i$ is stored and protected by a lower-rate balanced error-correcting codes (the misdecoding of $i$ may lead to catastrophic error propagation in the information word), we do not store $i$ in our construction. The main idea is that certain redundancy exists in the codewords of LDPC codes that enables us to locate $i$ or at last find a small set that includes $i$
with a very high probability, even some errors exist in the codewords. It is wasteful to store the value of $i$ with
a lower-rate balanced error-correcting code. As a result, our construction is more efficient than the recent construction proposed in \cite{Weber2012}.

\begin{figure}[!ht]
\centering
\includegraphics[width=3.4in]{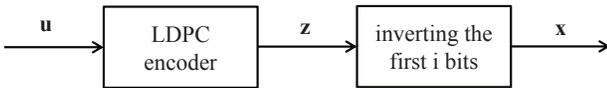}
\caption{Encoding of balanced LDPC codes.}
\label{fig_balanceencoding}
\end{figure}

Let $\mathbf{u}$ be the message to encode and its length is $k$, according to the description above,
the encoding procedure consists of two steps, as shown in Fig.~\ref{fig_balanceencoding}:
\begin{enumerate}
  \item Apply an $(n,k)$ LDPC code $\mathcal{L}$ to encode the message $\mathbf{u}$ into a codeword of length $n$, denoted by $\mathbf{z}=G\mathbf{u}$,
  where $G$ is the generator matrix of $\mathcal{L}$.
  \item Find the minimal integer $i$ in $\{0,1,...,n-1\}$ such that inverting the first $i$ bits of $\mathbf{z}$ results in
  a balanced word
  $$\mathbf{x}=\mathbf{z}+\mathbf{1}^i \mathbf{0}^{n-i},$$
  where $\mathbf{1}^i\mathbf{0}^{n-i}$ denotes a run of $i$ bits $\mathbf{1}$ and $n-i$ bits $\mathbf{0}$. Then we denote $\mathbf{x}$ as
  $\phi(\mathbf{z})$.
  This word $\mathbf{x}$ is
  a codeword of the resulting balanced LDPC code, denoted by $\mathcal{C}$.
\end{enumerate}

We see that a balanced LDPC code is constructed by simply balancing the codewords of a LDPC code, which is called the original LDPC code.
Based on the procedure above we can encode any message $\mathbf{u}$ of length $k$ into a balanced codeword $\mathbf{x}$ of length $n$.
The encoding procedure is very simple, but how to decode a received word? Now, we focus on the decoding of this balanced LDPC code.
Let $\mathbf{y}$ be an erroneous word received by the decoder, then the output of the maximum likelihood decoder is
$$\mathbf{\hat{x}}=\arg \min_{\mathbf{x}\in\mathcal{C}}  D(\mathbf{y}, \mathbf{x}),$$
where $D(\mathbf{y},\mathbf{x})$ is the distance between $\mathbf{y}$ and $\mathbf{x}$ depending on the channel, for instance, Hamming distance for binary symmetric channels.

The balanced code $\mathcal{C}$ is not a linear code, so the constraint $\mathbf{x} \in \mathcal{C}$ is not easy to deal with.
A simpler way is to think about the codeword $\mathbf{z}\in \mathcal{L}$ that corresponds to $\mathbf{x}$.  By inverting
the first $j$ bits of $\mathbf{y}$ with $0\leq j< n$, we can get a set of words $S_{\mathbf{y}}$ of size $n$, namely,
$$S_\mathbf{y}=\{\mathbf{y^{(0)}}, \mathbf{y^{(1)}}, ..., \mathbf{y^{(n-1)}}\},$$
in which
$$\mathbf{y^{(j)}}=\mathbf{y}+\mathbf{1}^j \mathbf{0}^{n-j},$$
for all $j\in \{0,1,2,...,n\}$. Then there exists an $i\in \{0,1,2,...,n-1\}$ such that
$$\mathbf{y^{(i)}}-\mathbf{z}=\mathbf{y}-\mathbf{x}.$$
The output of the maximum likelihood decoder is
$$(\mathbf{\hat{z}, \hat{i}})=\arg \min_{\mathbf{z'}\in \mathcal{L}, i'\in \{0,1,2...,n\}} D(\mathbf{y^{(i')}}, \mathbf{z'}),$$
subject to $i'$ is the minimum integer that makes $\mathbf{z'}+\mathbf{1}^{i'} \mathbf{0}^{n-i'}$
being balanced.

\begin{figure}[!t]
\centering
\includegraphics[width=3.4in]{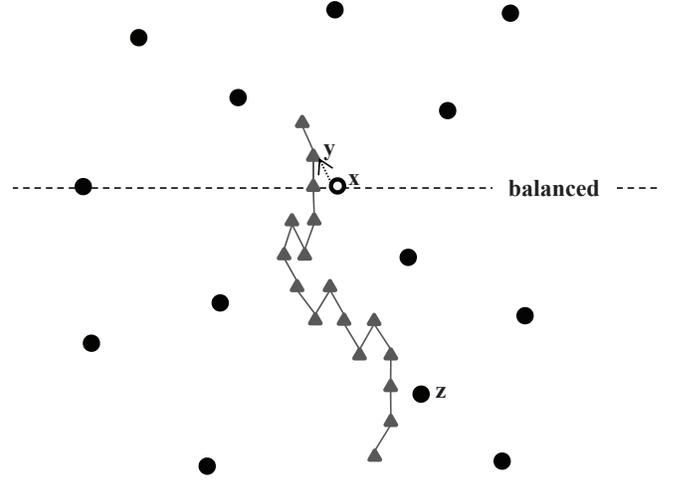}
\caption{Demonstration for the decoding of balanced LDPC codes.}
\label{fig_demonstration}
\end{figure}

If we ignore the constraint that $i$ has to be the minimum integer, then the output of the decoder is
the codeword in $\mathcal{L}$ that has the minimum distance to $S_\mathbf{y}$.
Fig.~\ref{fig_demonstration} provides
a simple demonstration, where the solid circles are for the codewords of the LPDC code $\mathcal{L}$,
the triangles are for the words in $S_{\mathbf{y}}$ that are connected by lines. Our goal is to find the solid circle that
is the closest one to the set of triangles. It is different from traditional decoding of linear codes whose goal is to find the closest codeword
to a single point.

\subsection{An Extreme Case}
\label{section_bal_LDPCextreme}

LDPC codes achieve near Shannon bound performances. A natural question is whether balanced LDPC codes
hold this property. Certain difficulties exist in proving it by following the method in \cite{Gallager63} (section 2 and section 3), since balanced LDPC codes are not linear codes and the distance distributions of balanced LDPC codes are not easy to characterize. Fortunately,
this statement looks correct because if the first $i$ bits of a codeword have been inverted (we assume that the interger
$i$ is unknown), then the codeword can be recovered with only little cost, i.e., a very small number of additional redundant bits.

Let us consider the ensemble of an $(n,a,b)$ parity-check matrix given by Gallager \cite{Gallager63}, which has $a$ ones in each column, $b$ ones in each row, and zeros elsewhere. According to this construction, the matrix is divided into $a$ submatrices, each containing a single $1$ in each column. All the submatrices are random column permutations of a matrix that has
a single one in each column and $b$ ones in each row. As a result, we have $(n,a,b)$ LDPC codes.

\begin{Theorem} Given a codeword $\mathbf{z}$ of an $(n,a,b)$ LDPC code, we get
$$\mathbf{x}=\mathbf{z}+\mathbf{1}^i \mathbf{0}^{n-i}$$
by inverting the first $i$ bits of $\mathbf{z}$ with $0\leq i<n$.
Let $P_e(\mathbf{x})$ be the error probability that $\mathbf{z}$ cannot be correctly recovered from $\mathbf{x}$ if $i$ is unknown.
As $n\rightarrow\infty$,
$$P_{e}(\mathbf{x})\rightarrow 0,$$
for any integers $a$ and $b$.
\end{Theorem}

\proof Let $H$ be the parity-check matrix of the LDPC code, and let $$\mathbf{y^{(j)}}= \mathbf{x}+\mathbf{1}^j \mathbf{0}^{n-j},$$
for all $j\in \{0,1,2,...,n-1\}$.

We can recover $\mathbf{z}$ from $\mathbf{x}$ if and only if
$$H \mathbf{y^{(j)}}\neq 0,$$
for all $j\neq i$ and $0\leq j\leq n-1$.

Hence,
$$P_e(\mathbf{x})= P(\exists j\neq i, s.t., H\mathbf{y^{(j)}}=0)$$
$$\leq \sum_{j\neq i} P(H\mathbf{y^{(j)}}=0).$$

Let us first consider the case of $j>i$. We have $H\mathbf{y^{(j)}}=0$ if and only if
$$H(\mathbf{y^{(j)}}+\mathbf{z})=0,$$
where $$\mathbf{y^{(j)}}+\mathbf{z}=\mathbf{0}^{i}\mathbf{1}^{j-i}\mathbf{0}^{n-j}.$$
So $H\mathbf{y^{(j)}}=0$ is equivalent to
$$H(\mathbf{0}^{i}\mathbf{1}^{j-i}\mathbf{0}^{n-j})=0.$$

As we described, $H$ is constructed by $a$ submatrices, namely, we can write $H$ as
$$H= \left(
      \begin{array}{c}
        H_1 \\
         H_2 \\
        \vdots \\
        H_a \\
      \end{array}
    \right).
$$

Let $H_s$ be one of the $a$ submatrices of $H$, then $H$ contains a single one in each columns and $b$ ones
in each row. And it satisfies
$$H_s(\mathbf{0}^{i}\mathbf{1}^{j-i}\mathbf{0}^{n-j})=0,$$
i.e., in each row of $H_s$, there are even number of ones from the $i+1$th column to the $j$th column.

According to the construction of $(n,a,b)$ LDPC codes,
$$P(H_s(\mathbf{0}^{i}\mathbf{1}^{j-i}\mathbf{0}^{n-j})=0)= P(H_s(\mathbf{1}^{j-i}\mathbf{0}^{n-j+i})=0).$$
So we can use $P(n,j-i)$ to denote $P(H_s(\mathbf{0}^{i}\mathbf{1}^{j-i}\mathbf{0}^{n-j})=0)$.

First, we consider the case that $b$ is even. In this case,
$$P(n,j-i)=P(n,n-j+i).$$
Hence, without loss of generality, we can assume that $j-i=d\leq \frac{n}{2}$.

It is easy to see that $P(n,j-i)>0$ only if $d$ is even.
Assume that the one in the first column of $H_s$ is in the $t$th row, and let $u$
be the number of ones in the $t$th row from the first $j-i$ columns. Then we can get
$$ P(n,d)= \sum_{u=2,4,...} \nchoosek{b}{u-1}(\frac{d-1}{n-1})^{u-1} $$
$$\times(\frac{n-d}{n-1})^{b-u} P(n-b, d-u),$$
where $P(n,d)=1$ if $n=d$ or $d=0$.

If $d<\log n$, then $P(n,d)=O(\frac{\log n}{n})$.

If $\log n \leq d\leq \frac{n}{2}$, then
$$\sum_{u=2,4,...} \nchoosek{b}{u-1}(\frac{d-1}{n-1})^{u-1} (\frac{n-d}{n-1})^{b-u}\leq \frac{b-1}{b}.$$
Iteratively, we can prove that
$$P(n,d)=O((\frac{b-1}{b})^{\frac{\log n}{2b}}).$$

Similar as above, when $j<i$, we can get $$P(H\mathbf{y^{(j)}}=0)\leq P(n,i-j).$$

Finally, we have
$$P_e(\mathbf{x})\leq \sum_{s=1}^{n-1-i}P(n,s)+\sum_{s=1}^{i}P(n,s)=O(\frac{\log n}{n}).$$

So if $b$ is even, as $n\rightarrow\infty$, $P_e(\mathbf{x})\rightarrow0$.

If $b$ is odd, in each row, there exists at least one $1$ in the last $n-j+i$ elements. As a result,
$n-j+i\geq \frac{n}{b}$. Using a same idea as above,
we can also prove that as $n\rightarrow\infty$, $P_e(\mathbf{x})\rightarrow 0$.

So the statement in the theorem is true for any rate $R=\frac{b-a}{b}<1$.
This completes the proof.\hfill\QED

The above theorem considers an extreme case that if the codeword of a balanced LDPC code does not have
errors, then we can recover the original message with little cost of redundancy. It implies that balanced
LDPC codes may achieve almost the same rates as the original unbalanced LDPC codes. In the following subsections,
we discuss some decoding techniques for binary erasure channels and binary symmetric channels. Simulation results
on these channels support the above statement.

\subsection{Decoding for Erasure Channels}
\label{section_bal_LDPCerasure}

In this subsection, we consider binary erasure channels (BEC), where a bit ($0$ or $1$) is either successfully received or it is deleted, denoted by ``$?$".
Let $\mathbf{y}\in\{0,1,?\}^n$ be a word
received by a decoder after transmitting a codeword $\mathbf{x}\in \mathcal{C}$ over a BEC.
Then the key of decoding $\mathbf{y}$ is to determine the value of the integer $i$ such that $\mathbf{x}$ can be obtained
by inverting the first $i$ bits of a codeword in $\mathcal{L}$.

A simple idea is to search all the possible values of $i$, i.e., we decode all the possible words
$\mathbf{y^{(0)}}, \mathbf{y^{(1)}}, ..., \mathbf{y^{(n-1)}}$ separately and select the best
 resulting codeword that satisfies all the constraints as the final output. This idea is straightforward, but the computational complexity of the
decoding increases by a factor of $n$, which is not acceptable for most practical applications.

Our observation is that we might be able to determine the value of $i$ or at least find a feasible set that includes $i$, based on
the unerased bits in $\mathbf{y}$. For example, given $\x\in \mathcal{L}$, assume that one parity-check constraint is
$$x_{i_1}+x_{i_2}+...+x_{i_4}=0.$$
If all $y_{i_1}, y_{i_2}, ..., y_{i_4}$ are observed (not erased), then we can have the following statement about $i$:

(1) If $y_{i_1}+y_{i_2}+...+y_{i_4}=0$, then
$$i\in [0,i_1)\bigcup [i_2, i_3) \bigcup [i_4,n].$$

(2) If $y_{i_1}+y_{i_2}+...+y_{i_4}=1$, then
$$i\in [i_1,i_2)\bigcup[i_3,i_4).$$

\begin{figure}[!t]
\centering
\includegraphics[width=3.4in]{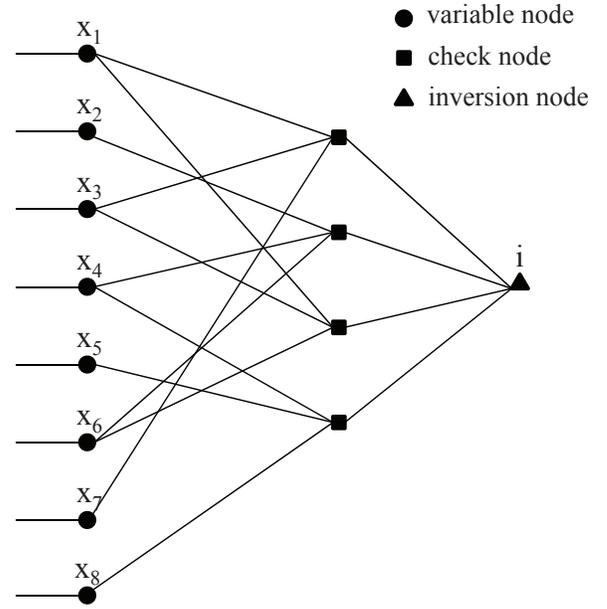}
\caption{Graph for balanced LDPC codes.}
\label{fig_decodinggraph}
\end{figure}

By combining this observation with the message-passing algorithm, we get a decoding algorithm for balanced LDPC codes under BEC. Similar as the original LDPC code, we present a balanced LDPC code as a sparse bipartite graph with
$n$ variable nodes and $r$ check nodes, as shown in Fig.~\ref{fig_decodinggraph}. Additionally,
we add an inversion node for representing the value or the feasible set of $i$.  Let us
describe a modified message-passing algorithm on this graph. In each round of the algorithm,
messages are passed from variable nodes and inversion nodes to check nodes, and then
from check nodes back to variable nodes and inversion nodes.

We use $\mathcal{I}$ denote the feasible set consisting of all possible values for the integer $i$, called inversion set. At the first round, we initialize the $j$th variable node $y_j\in \{0,1,?\}$ and
initialize the inversion set as $\mathcal{I}=[0,n]$. Then we pass message and update the graph iteratively. In each round, we do the following operations.

(1) For each variable node $\mathbf{v}$, if its value $x_v$ is in $\{0,1\}$, it sends $x_v$ to all its check neighbors.
If $x_v=?$ and any incoming message $u$ is $0$ or $1$, it updates $x_v$ as $u$ and sends $u$ to all its check neighbors.
If $x_v=?$ and all the incoming messages are $?$, it sends $?$ to all its check neighbors.

(2) For each check node $\mathbf{c}$, assume the messages from its variable neighbors are $x_{i_1}, x_{i_2}, ..., x_{i_b}$, where
$i_1, i_2, ..., i_b$ are the indices of these variable nodes s.t. $i_1<i_2<...<i_b$. Then
we define $$S_c^0=[0,i_1)\bigcup [i_2,i_3) \bigcup ...,$$ $$S_c^1=[i_1,i_2)\bigcup[ i_3,i_4)\bigcup ....$$
If all the incoming messages are in $\{0,1\}$, then we update $\mathcal{I}$ in the following way: If
$x_{i_1}+x_{i_2}+...+x_{i_b}=0$, we update $\mathcal{I}$ as $\mathcal{I}\bigcap S_c^0$; otherwise, we update
$\mathcal{I}$ as $\mathcal{I}\bigcap S_c^1$. In this case, this check node $\mathbf{c}$ is no longer useful, so
we can remove this check node from the graph.

(3) For each check node $\mathbf{c}$, if there are exactly one incoming message from its variable neighbor which is $x_j=?$ and
all other incoming messages are in $\{0,1\}$, we check whether $\mathcal{I}\subseteq S_c^0$ or $\mathcal{I}\subseteq  S_c^1$.
If $\mathcal{I} \subseteq S_c^0$, then the check node sends the XOR of the other incoming messages except $?$ to $x_j$.
If $\mathcal{I} \subseteq S_c^1$, then the check node sends the XOR of the other incoming messages except $?$ plus one to $x_j$.
In this case, the check node $\mathbf{c}$ is also no longer useful, so we can remove this check node from the graph.

The procedure above continues until all erasures are filled in, or no erasures are filled in the current iteration.
Different from the message-passing decoding algorithm for LDPC codes, where
in each iteration both variable nodes and check nodes are processed only once, here, we process variable nodes once but
check nodes twice in each iteration. If all erasures are filled in, $\mathbf{x}$ is the binary vector labeled
on the variable nodes. In this case, if $|\mathcal{I}|=1$, then $i$ is the only element in $\mathcal{I}$, and
we can get $\mathbf{z}\in \mathcal{L}$ by calculating
$$\mathbf{z}=\mathbf{x}+\mathbf{1}^i \mathbf{0}^{n-i}.$$

If there are still some unknown erasures, we enumerate all the possible values in $\mathcal{I}$ for the integer $i$.
Usually, $|\mathcal{I}|$ is small. For a specific $i$, it leads to a feasible solution $\mathbf{z}$ if

(1) Given $\mathcal{I}=\{i\}$, with the message-passing procedure above, all the erasures can be filled in.

(2) $\mathbf{x}$ is balanced, namely, the numbers of ones and zeros are equal for the variable nodes.

(3) Let $\mathbf{z}=\mathbf{x}+\mathbf{1}^i\mathbf{0}^{n-i}$. Then $i$ is the minimal integer in $\{0,1,2,...,n\}$ subject to $\mathbf{z}+\mathbf{1}^i\mathbf{0}^{n-i}$ is balanced.

We say that a word $\mathbf{y}$ with erasures is uniquely decodable if and only if there exists $i\in \mathcal{I}$ that leads to a feasible solution, and for all such integers $i$ they result in the unique solution $\mathbf{z}\in \mathcal{L}$. The following simple example
is provided for the purpose of demonstrating the decoding process.

\begin{Example} Based on Fig.~\ref{fig_decodinggraph}, we have a codeword  $\mathbf{x}=01111000$, which is transmitted over
an erasure channel. We assume that the received word is $\mathbf{y}=011110??$.

In the first round of the decoding, we have
$$\mathbf{x^{(1)}}=011110??, \mathcal{I}=[0,8].$$

Considering the $2$nd check node, we can update $\mathcal{I}$ as
$$\mathcal{I}=\{0,1, 4, 5 \}.$$

Considering the $3$nd check node, we can continue updating $\mathcal{I}$ as
$$\mathcal{I}=\mathcal{I}\bigcap \{1,2,6,7,8\}=\{1\}.$$

Based on (3), we can fill $0, 0$ for the $7$th and $8$th variable nodes.
Finally, we get $\mathbf{z}=11111000$ and $i=1$.
\end{Example}

\begin{figure}[!t]
\centering
\includegraphics[width=3.8in]{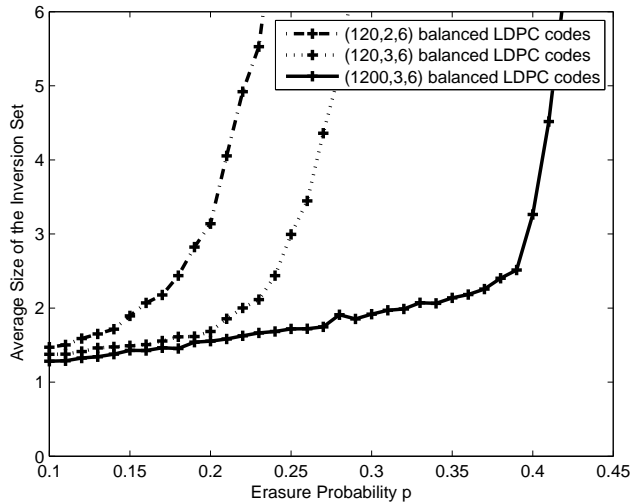}
\caption{The average size of the inversion set $\mathcal{I}$ after iterations in the message-passing algorithm for decoding balanced LDPC codes.}
\label{fig_inversionset}
\end{figure}

Regarding to the decoding algorithm described above, there are two important issues that need to consider, including
the decoding complexity of the algorithm and its performance. First, the decoding complexity of the
algorithm strongly depends on the size of $\mathcal{I}$ when it finishes iterations. Fig.~\ref{fig_inversionset}
simulates the average size of the inversion set $\mathcal{I}$ for decoding three balanced LDPC codes.
It shows that when the crossover probability is lower than a threshold, the size of $\mathcal{I}$ is smaller than
a constant with a very high probability. In this case, the decoding complexity of the balanced LDPC code
is very close to the decoding complexity of the original unbalanced LDPC code.

\begin{figure}[!t]
\centering
\includegraphics[width=3.8in]{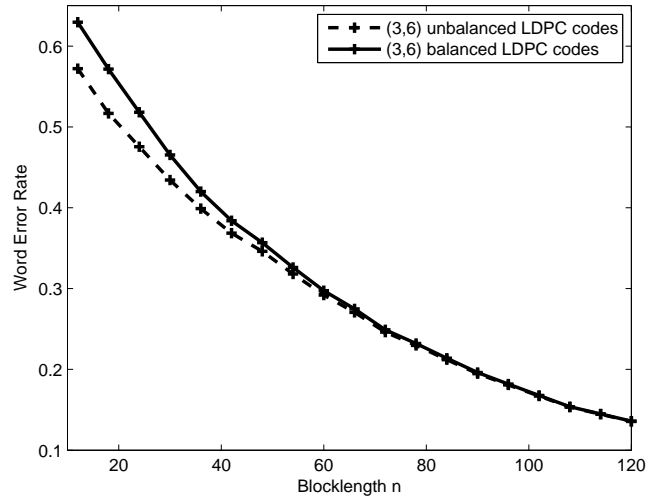}
\caption{Word error rate of balanced LDPC codes and unbalanced LDPC codes when the erasure probability $p=0.35$.}
\label{fig_erasure_error}
\end{figure}

Another issue is about the performance of the decoding algorithm for balanced LDPC codes. In particular, we want to figure out
the cost of additional redundancy in correcting the inversion of the first $i$ bits when $i$ is unknown.
In Fig.~\ref{fig_erasure_error}, it presents the word error rate of balanced LDPC codes and the corresponding original unbalanced LDPC codes
for different block lengths. It is interesting to see that as the block length increases, the balanced LDPC codes and the original unbalanced LDPC
codes have almost the same performance, that is, the cost of correcting the inversion of the first $i$ bits is ignorable.

\subsection{Decoding for Symmetric Channels}
\label{section_bal_LDPCsymmetric}

In this subsection, we study and analyze the decoding of balanced LDPC codes for symmetric channels, including binary symmetric channels (BSC) and AWGN (Additive White Gaussian Noise) channels. Different from binary erasure channels (BEC), here we are not able to determine a small set that definitely includes the integer $i$. Instead,
we want to figure out the most possible values for $i$. Before presenting our decoding algorithm,
we first introduce belief propagation algorithm for decoding LDPC codes.

Belief propagation \cite{McEliece98}, where messages are passed iteratively across a factor graph, has been widely studied and recommended for the decoding of LDPC codes. In each iteration, each variable node passes messages (probabilities) to all the adjacent check nodes and
then each check node passes messages (beliefs) to all the adjacent variable nodes. Specifically, let $\mathbf{m}_{vc}^{(\ell)}$ be the message passed from a variable node $\mathbf{v}$ to a check node $\mathbf{c}$ at
the $\ell$th round of the algorithm, and let $\mathbf{m}_{cv}^{(\ell)}$ be the message from a check node $\mathbf{c}$ to a variable node $\mathbf{v}$. At the first round, $\mathbf{m}_{vc}^{(0)}$ is the log-likelihood of
the node $\mathbf{v}$ conditioned on its observed value, i.e.,
$\log \frac{P(y|x=0)}{P(y|x=1)}$ for variable $x$ and its observation $y$. This value is denoted by $\mathbf{m}_v$. Then the iterative update procedures can be described by the following equations
$$\mathbf{m}_{vc}^{(\ell)}=\left\{\begin{array}{cc}
                                    \mathbf{m}_v &  \ell=0,\\
                                    \mathbf{m}_v + \sum_{c'\in N(v)/c} \mathbf{m}_{c'v}^{(\ell-1)} \quad&  \ell\geq 1,
                                  \end{array}
\right.$$
$$\mathbf{m}_{cv}^{(\ell)}=2\tanh^{-1}(\prod_{v'\in N(c)/v} \tanh(\frac{\mathbf{m}_{v'c}^{(\ell)}}{2})),$$
where $N(v)$ is the set of check nodes that connect to variable node $\mathbf{v}$ and $N(c)$
is the set of variable nodes that connect to check node $\mathbf{c}$. In practice, the belief-propagation algorithm stops after a certain number of iterations or until the passed likelihoods are close to certainty. Typically, for a BSC with crossover probability $p$, the log-likelihood
$\mathbf{m}_{v}$ for each variable node $\mathbf{v}$ is a constant depending on $p$. Let $x$ be
the variable on $\mathbf{v}$ and let $y$ be its observation, then
$$\mathbf{m}_{v}=\left\{\begin{array}{cc}
                          \log \frac{1-p}{p} & \textrm{ if }y=0,\\
                          -\log \frac{1-p}{p} & \textrm{ if }y=1. \\
                        \end{array}
\right.$$

Let us consider the decoding of balanced LDPC codes. Assume $\mathbf{x}\in \mathcal{C}$ is a codeword of a balanced LDPC code,
obtained by inverting the first $i$ bits of a codeword $\mathbf{z}$ in a LDPC code $\mathcal{L}$.
The erroneous word received by the decoder is $\mathbf{y}\in \mathcal{Y}^n$ for an alphabet $\mathcal{Y}$.
For example, $\mathcal{Y}=\{0,1\}$ for BSC channels, and $\mathcal{Y}=\mathbb{R}$ for AWGN channels.
Here, we consider a symmetric channel, i.e., a channel for which
there exists a permutation $\pi$ of the output alphabet $\mathcal{Y}$ such that (1)
$\pi^{-1}=\pi$, and (2) $P(y|1)=P(\pi(y)|0)$ for all $y\in \mathcal{Y}$, where
$P(y|x)$ is the probability of observing $y$ when the input bit is $x$.

The biggest challenge of decoding a received word $\mathbf{y}\in \mathcal{Y}^n$ is
lacking of the location information about where the inversion happens, i.e., the integer $i$.
We let $$\mathbf{y^{(i)}}=\pi(y_1)\pi(y_2)...\pi(y_i)y_{i+1}...y_n,$$
for all $i\in \{0,1,2,...,n-1\}$.
A simple idea is to search all the possibilities for the integer $i$ from $0$ to $n-1$, i.e,  decoding
all the words
$$ \mathbf{y^{(0)}}, \mathbf{y^{(1)}}, ..., \mathbf{y^{(n-1)}}$$
separately. Assume their decoding outputs based on belief propagation are
$$\mathbf{\hat{z}^{(0)}}, \mathbf{\hat{z}^{(1)}}, ... \mathbf{\hat{z}^{(n)}},$$
then the final output of the decoder is $\mathbf{\hat{z}}=\mathbf{\hat{z}^{(j)}}$ such that
$P(\mathbf{y^{(j)}}|\mathbf{\hat{z}^{(j)}})$ is maximized. The drawback of this method is
its high computational complexity, which is about $n$ times the complexity of decoding the original unbalanced LDPC code.
To reduce computational complexity, we
want to estimate the value of $i$ in a simpler and faster way, even sacrificing a little bit of performance on bit error rate.

The idea is that when we are using belief propagation to decode a group of words $\mathbf{y^{(0)}}, \mathbf{y^{(1)}}, ..., \mathbf{y^{(n-1)}}$,
some information can be used to roughly compare their goodness, namely, their distances to the nearest codewords.
To find such information, given each word $\mathbf{y^{(i)}}$ (here, we denote it as $\mathbf{y}$ for simplicity),
we run belief propagation for $\ell$ rounds (iterations), where $\ell$ is very small, e.g., $\ell=2$.
There are several ways of estimating the goodness of $\mathbf{y}$, and we introduce one of them as follows.

Given a word $\mathbf{y}$, we define
$$\lambda(\mathbf{y},\ell)= \sum_{c\in C} \prod_{v\in N(c)} \tanh(\mathbf{m}_{vc}^{(\ell)}/2),$$
where $C$ is the set of all the variable nodes, $N(c)$
is the set of neighbors of a check node $\mathbf{c}$, and $\mathbf{m}_{vc}^{(\ell)}$ is the message passed from a variable node $\mathbf{v}$ to a check node $\mathbf{c}$ at
the $\ell$th round of the belief-propagation algorithm.
Roughly, $\lambda(\mathbf{y},\ell)$ is a measurement of the number of correct parity checks for the current assignment
in belief propagation (after $\ell-1$ iterations). For instance,
$$\lambda(\mathbf{y},\ell=1)= \alpha(r-2|H\mathbf{y}|),$$
for a binary symmetric channel. In this expression, $\alpha$ is a constant, $r=n-k$ is the number of redundancies, and $|H\mathbf{y}|$ is the number of ones in $H\mathbf{y}$, i.e., the number of unsatisfied parity checks.

Generally, the bigger $\lambda(\mathbf{y^{(j)}},\ell)$ is, the more likely $j=i$ is.
So we can get the most likely $i$ by calculating
$$\hat{i}=\arg\max_{j=0}^{n-1} \lambda(\mathbf{y^{(j)}},\ell).$$
Then we decode $\mathbf{y^{(\hat{i})}}$ as the final output. However, the procedure requires to calculate  $\lambda(\mathbf{y^{(j)}},\ell)$ with $0\leq j\leq n-1$.
The following theorem shows that the task of computing all $\lambda(\mathbf{y^{(j)}},\ell)$ with $0\leq j\leq n-1$ can be
finished in linear time if $\ell$ is a small constant.

\begin{Theorem}
The task of computing all $\lambda(\mathbf{y^{(j)}},\ell)$ with $0\leq j\leq n-1$ can be
finished in linear time if $\ell$ is a small constant.
\end{Theorem}

\proof
First, we calculate $\lambda(\mathbf{y^{(0)}},\ell)$. Based on the belief-propagation algorithm described above,
it can be finished in $O(n)$ time. In this step, we save all the messages including $\mathbf{m}_v$, $\mathbf{m}_{cv}^{(l)}$,
$\mathbf{m}_{vc}^{(l)}$ for all $c\in C, v\in V$ and $1\leq l\leq \ell$.

When we calculate $\lambda(\mathbf{y^{(1)}}, \ell)$, the only change on the inputs is $\mathbf{m}_{v_1}$, where $v_1$ is the first variable node (the sign of $\mathbf{m}_{v_1}$ is flipped).
As a result, we do not have to calculate all $\mathbf{m}_v$, $\mathbf{m}_{cv}^{(l)}$,
$\mathbf{m}_{vc}^{(l)}$ for all $c\in C, v\in V$ and $1\leq l\leq \ell$. Instead, we only need to update those messages that are related with
$\mathbf{m}_{v_1}$. It needs to be noted that the number of messages related to $\mathbf{m}_{v_1}$ has an exponential dependence on $\ell$, so
the value of $\ell$ should be small. In this case, based on the calculation of $\lambda(\mathbf{y}^{(0)},\ell)$,
$\lambda(\mathbf{y^{(1)}}, \ell)$ can be calculated in a constant time. Similarly, each of $\lambda(\mathbf{y^{(j)}},\ell)$ with $2\leq j\leq n-1$
can be obtained iteratively in a constant time.

Based on the process above, we can compute
all $\lambda(\mathbf{y^{(j)}},\ell)$ with $0\leq j\leq n-1$ in $O(n)$ time.
\hfill\QED

To increase the success rate of decoding, we can also create a set of most likely values for $i$, denoted by $\mathcal{I}_c$.
$\mathcal{I}_{c}$ consists of at most $c$ local maximums with the highest values of $\lambda(\mathbf{y^{(i)}},\ell)$. Here, we say
that $j\in \{0,1, 2,3,...,n-1\}$ is a local maximum if and only if
$$\lambda(\mathbf{y^{(j)}},\ell)> \lambda(\mathbf{y^{(j-1)}},\ell), \lambda(\mathbf{y^{(j)}},\ell)\geq  \lambda(\mathbf{y^{(j+1)}},\ell).$$
Note that $\mathcal{I}_1=\{\hat{i}\}$, where $\hat{i}$ is the global maximum as defined above. If $c>1$, for all $j\in \mathcal{I}_c$, we decode
$\mathbf{y^{(j)}}$ separately and choose the output with the maximum likelihood as the final output of the decoder.
It is easy to see that the the above modified belief-propagation algorithm for balanced LDPC codes has asymptotically the same decoding complexity as the belief-propagation algorithm for LDPC codes, that is, $O(n\log n)$.

\begin{figure*}[!t]
\centering
\includegraphics[width=5in]{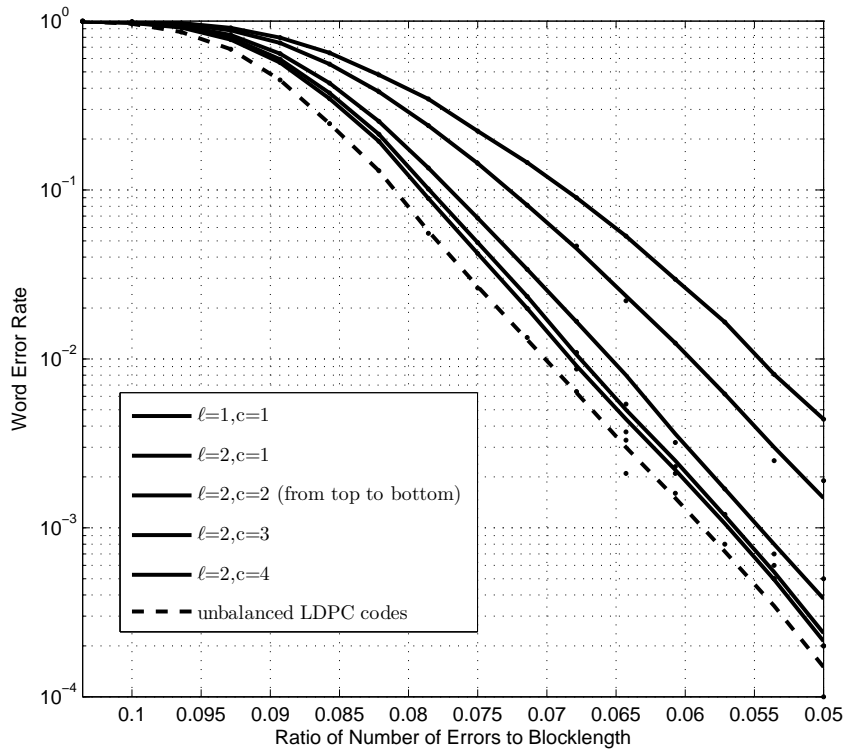}
\caption{World error rate of $(280,4,7)$ LDPC codes with maximal $50$ iterations.}
\label{fig_errorword}
\end{figure*}

In Fig.~\ref{fig_errorword}, it shows the performance of the above algorithm for decoding
balanced LDPC codes under BSC and the performance of belief propagation algorithm for the original LDPC codes. From which,
we see that when $\ell=2$ and $c=4$, the performance gap between balanced $(280,4,7)$ LDPC code and unbalanced $(280,4,7)$ LDPC code is very small.
This comparison implies that the
cost of correcting the inversion of the first $i$ bits (when $i$ is unknown) is small for LDPC codes.

Let us go back the scheme of balanced modulation. The following examples give
the log-likelihood of each variable node when the reading process is based on hard decision
and soft decision, respectively. Based on them, we can apply the modified propagation algorithm
in balanced modulation.

\begin{Example}
If the reading process is based on hard decision, then
it results in a binary symmetric channel with crossover probability $p$. In this case, let $y$ be the
observation on a variable node $\mathbf{v}$, the log-likelihood for $\mathbf{v}$ is
$$\mathbf{m}_{v}=\left\{\begin{array}{cc}
                          \log \frac{1-p}{p} & \textrm{ if }y=0, \\
                          -\log \frac{1-p}{p} & \textrm{ if }y=1. \\
                        \end{array}
\right.$$
\end{Example}

\begin{Example}
If the reading process is based on soft decision, then we can approximate
cell-level distributions by Gaussian distributions, which are characterized by $4$ parameters $u_0, \sigma_0, u_1, \sigma_1$.
These parameters can be obtained based on the cell-level vector $\mathbf{y}=\mathbf{c}$, following the steps in Subsection \ref{subsection_soft}.
In this case, if the input of the decoder is $\mathbf{y}$, then the log-likelihood of the $i$th variable node $\mathbf{v}$ is
$$\mathbf{m}_{v}=\lambda_i=\frac{\log\frac{1}{\sigma_0}-\frac{(c_i-u_0)^2}{2\sigma_0^2}}{\log\frac{1}{\sigma_1}-\frac{(c_i-u_1)^2}{2\sigma_1^2}}$$
where $c_i$ is the current level of the $i$th cell.
If the input of the decoder is $\mathbf{y^{(i)}}$ (we don't have to care about its exact value), then the log-likelihood of the $i$th variable node $\mathbf{v}$ is
$$\mathbf{m}_{v}=\left\{\begin{array}{cc}
                          \lambda_i& \textrm{ if } i>j, \\
                          -\lambda_i & \textrm{ if } i\leq j, \\
                        \end{array}
\right.,$$
for all $0\leq i<n$.
\end{Example}

\section{Partial-Balanced Modulation}

\label{section_bal_variant}

Constructing balanced error-correcting codes is more difficult than constructing normal error-correcting codes. A question is: is it possible
to design some schemes that achieve similar performances with balanced modulation and have simple error-correcting code constructions?
With this motivation, we propose a variant of balanced modulation, called partial-balanced modulation.
The main idea is to construct an error-correcting code whose codewords are partially balanced, namely, only a certain segment of each codeword is balanced. When reading information from a block, we adjust the reading threshold to make this segment of the resulting word being balanced or being approximately balanced.

\begin{figure}[!t]
\centering
\includegraphics[width=3.2in]{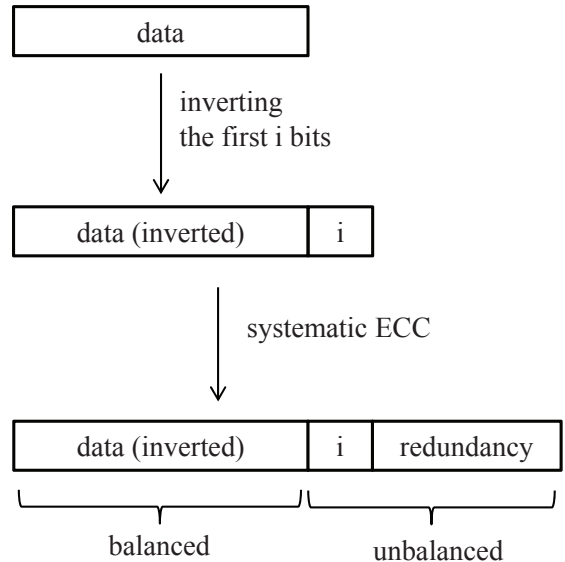}
\caption{Partial balanced code.}
\label{fig_particalbalance}
\end{figure}

One way of constructing partial-balanced error-correcting codes is shown in Fig.~\ref{fig_particalbalance}.
Given an information vector $\mathbf{u}$ of $k$ bits ($k$ is even), according to Knuth's observation \cite{Knuth86},
there exists an integer $i$ with $0\leq i< k$ such that
inverting the first $i$ bits of $\mathbf{u}$ results in a balanced word $\mathbf{\widetilde{u}}$. Since our goal is to construct a codeword that is
partially balanced, it is not necessary to present $i$ in a balanced form. Now, we use $\mathbf{i}$ denote
the binary representation of length $\lceil\log_2 k\rceil$ for $i$.
To further correct potential errors,
we consider $[\mathbf{\widetilde{u}}, \mathbf{i}]$ as the information part and add extra parity-check bits by applying a systematic error-correcting code, like BCH code, Reed-Solomon code, etc. As a result, we obtain a codeword $\mathbf{x}=[\mathbf{\widetilde{u}}, \mathbf{i}, \mathbf{r}]$ where
$\mathbf{r}$ is the redundancy part. In this codeword, $\mathbf{\widetilde{u}}$ is balanced, $[\mathbf{i}, \mathbf{r}]$ is not balanced.

Note that in most data-storage applications, the bit error rate of a block is usually very small. The application of modulation schemes can
further reduce the bit error rate. Hence, the number of errors in real applications is usually much smaller than the block length.
In this case, the total length of $[\mathbf{i}, \mathbf{r}]$ is smaller or much smaller than the code dimension $k$. As the block length $n$ becomes large, like one thousand, the reading threshold determined by partial-balanced modulation is almost the same as
the one determined by balanced modulation.  One assumption that we made is that all the cells in the same block have
similar noise properties. To make this assumption being sound, we can reorder the bits in $\mathbf{x}=[\mathbf{\widetilde{u}}, \mathbf{i}, \mathbf{r}]$ such that the $k$ cells of storing $\mathbf{\widetilde{u}}$ is (approximately) randomly distributed among all the $n$ cells.
Compared to balanced modulation, partial-balanced modulation can achieve almost the same performance, and its
code construction is much easier (the constraints on the codewords are relaxed). In the following two examples, it
compares the partial-balanced modulation scheme with the traditional one based on a fixed threshold.

\begin{Example} Let us consider a nonvolatile memory with block length $n=255$.
To guarantee the data reliability, each block has to correct $18$ errors if the reading process is based on
a fixed reading threshold. Assume $(255,131)$ primitive BCH code is applied for correcting errors, then the data rate (defined by the ratio between the number of
available information bits and the block length) is
$$\frac{131}{255}=0.5137.$$
\end{Example}

\begin{Example} For the block discussed in the previous example, we assume that it only needs to correct $8$ errors based on partial-balanced modulation. In this case,
we can apply $(255,191)$ primitive BCH code for correcting errors, and the data rate is
$$\frac{191-8}{255}=0.7176,$$
which is much higher than the one obtained in the previous example.
\end{Example}

The reading/decoding process of partial-balanced modulation is straightforward. First, the reading threshold $v_b$ is adjusted such that
among the cells corresponding to $\mathbf{u}$ there are $k/2$ cells or approximately $k/2$ cells
with higher levels than $v_b$. Based on this reading threshold $v_b$, the whole block is read as a binary word $\mathbf{y}$, which
can be further decoded as $[\mathbf{\widetilde{u}}, \mathbf{i}]$ if the total number of errors is well bounded.
Then we obtain the original message $\mathbf{u}$ by inverting the first $i$ bits of $\mathbf{\widetilde{u}}$.

\section{Balanced Codes for Multi-Level Cells}

\label{section_bal_multicell}

In order to maximize the storage capacity of nonvolatile memories,
multi-level cells (MLCs) are used, where
a cell of $q$ discrete levels can store $\log_2 q$ bits \cite{Brewer2008}. Flash memories with 4 and 8 levels have been used in
products, and MLCs with $16$ levels have been demonstrated
in prototypes. For PCMs, cells with $4$ or more levels have
been in development.

The idea of balanced modulation and partial-balanced modulation can be extended to multi-level cells. For instance,
if each cell has $4$ levels, we can construct a balanced code in which each codeword has the same number of $0$s, $1$s, $2$s, and $3$s.
When reading data from the block, we adjust three reading thresholds such that
the resulting word also has the same number of $0$s, $1$s, $2$s, and $3$s.
The key question is how to construct balanced codes or partial-balanced codes for an alphabet size $q>2$.

\subsection{Construction based on Rank}

A simple approach of constructing  balanced codes for a nonbinary case is to consider the message
as the rank of its codeword among all its permutations,  based on the lexicography order. If
the message is $\mathbf{u}\in \{0,1\}^k$, then the codeword length $n$ is the minimum integer such that
$n=qm$ and
$\genfrac(){0cm}{0}{qm}{m \quad m \quad ... \quad m}>2^{k}.$  The following examples are provided for demonstrating the
encoding and decoding processes.

\begin{Example}
Assume the message is $\mathbf{u}=1010010010$ of length $10$ and $q=3$.
Since $\genfrac(){0cm}{0}{9}{3 \quad 3 \quad 3}>2^{10}$, we can convert
$\mathbf{u}$ to a balanced word $\mathbf{x}$ of length $9$ and alphabet size $q=3$. Let $S$ denote the set that consists of
all the balanced words of length $9$ and alphabet size $q=3$. To map $\mathbf{u}$ into a word in $S$, we write $\mathbf{u}$
into the decimal form $r=658$ and let $r$ be the rank of $\mathbf{x}$ in $S$ based on the lexicographical order.

Let us consider the first symbol of $\mathbf{x}$. In $S$, there are totally $\genfrac(){0cm}{0}{8}{2 \quad 3 \quad 3}=560$ sequences starting
with $0$, or $1$, or $2$.  Since $560\leq r <560+560$, the first symbol in $\mathbf{x}$ would be $1$, then we update $r$ as $r-560=98$, which
is the rank of $\mathbf{x}$ among all the sequences starting with $1$.

Let us consider the second symbol of $\mathbf{x}$. There are totally $\genfrac(){0cm}{0}{8}{2 \quad 2 \quad 3}$ sequences
starting with $10$, and it is larger than $r$, so the second symbol of $\mathbf{x}$ is $0$.

Repeating this process, we can convert $\mathbf{u}$ into a balanced word $\mathbf{x}=101202102$.
\end{Example}

\begin{Example}
We use the same notations as the above example. Given $\mathbf{x}=101202102$, it is easy to calculate its rank in $S$ based on the lexicographical order (via enumerative source
coding \cite{Cover73}). It is
\begin{eqnarray*}
  r &=& \genfrac(){0cm}{0}{8}{2 \quad 3 \quad 3}+\genfrac(){0cm}{0}{6}{1 \quad 2 \quad 3}+\genfrac(){0cm}{0}{5}{1 \quad 1 \quad 3}\\
  &&+\genfrac(){0cm}{0}{5}{2 \quad 0 \quad 3} +\genfrac(){0cm}{0}{3}{0 \quad 1 \quad 2}+\genfrac(){0cm}{0}{3}{1 \quad 0 \quad 2}\\
  &&+\genfrac(){0cm}{0}{2}{0 \quad 1 \quad 1} \\
   &=& 656,
\end{eqnarray*}
where $\genfrac(){0cm}{0}{8}{2 \quad 3 \quad 3}$ is the number of $\mathbf{x}$'s permutations starting with $0$,
$\genfrac(){0cm}{0}{6}{1 \quad 2 \quad 3}$ is the number of $\mathbf{x}'$ permutations starting with $100$, ...

Then from $r$, we can get its binary representation $\mathbf{u}=1010010010$.  In \cite{Ryabko2000}, Ryabko and Matchikina showed that if the length of $\mathbf{x}$ is $n$, then we can get the message $\mathbf{u}$ in $O(n \log^3 n \log\log n)$ time.
\end{Example}

The above approach is simple and information efficient, but the encoding is not computationally fast.

\subsection{Generalizing Knuth's Construction}

An alternative approach is to generalize Knuth's idea to the nonbinary case due to its operational simplicity.
Generally, assume that we are provided a word $\mathbf{u}\in G_q^k$ with $G_q=\{0,1,2,...,q-1\}$ and $k=qm$,
our goal is to generalize Knuth's idea to make $\mathbf{u}$ being balanced.

Let us consider a simple case, $q=4$. Given a word $\mathbf{u}\in G_4^k$, we let $n_i$ with $0\leq i\leq 3$ denote the number of $i$s in $\mathbf{u}$. To balance all the cell levels, we first balance the total number of $0$s and $1$s, such that $n_0+n_1=2m$. It also results in $n_2+n_3=2m$.
To do this, we can treat $0$ and $1$ as an identical state and treat $2$ and $3$ as another identical state. Based on
Knuth's idea, there always exists an integer $i$ such that by operating on the first $i$ symbols ($0\rightarrow 2$, $1\rightarrow 3$, $2\rightarrow 0$,
$3\rightarrow 1$) it yields $n_0+n_1=2m$.
We then consider the subsequence consisting of $0$s and $1$s, whose length is $2m$. By applying Knuth's idea, we
can make this subsequence being balanced. Similarly, we can also balance the subsequence consisting of $2$s and $3$s.
Consequently, we convert any word in $G_4^k$ into a balanced word. In order to decode
this word, three additional integers of length at most $\lceil\log k\rceil$ need to be stored, indicating the locations of having operations. The following example is constructed
for the purpose of demonstrating this procedure.

\begin{Example} Assume $\mathbf{u}=0110230210110003$, we convert it into a balanced word with the following steps:

(1) By operating the first $4$ symbols in $\mathbf{u}$, it yields $2332230210110003$, where $n_0+n_1=8$.

(2) Considering the subsequence of $0$s and $1$s, i.e., the underlined part in $233223\underline{0}2\underline{1011000}3$. By operating the first bit of this subsequence $(0\rightarrow 1, 1\rightarrow 0)$, it yields $233223\underline{1}2\underline{1011000}3$, where $n_0=n_1=4$.

(3) Considering the subsequence of $0$s and $1$s, i.e., the underlined part in $\underline{233223}1\underline{2}1011000\underline{3}$. By operating the first $0$ bit of this subsequence $(2\rightarrow 3, 3\rightarrow 2)$, it yields $\underline{233223}1\underline{2}1011000\underline{3}$,
which is balanced.

To recover $0110230210110003$ from $2332231210110003$ (the inverse process), we need to record the three integers $[4,1,0]$ whose binary lengths
are $[\log_2 16, \log_2 8, \log_2 8]$.
\end{Example}

It can be observed that the procedure above can be easily generalized for any $q=2^a$ with $a\geq 2$. If $m=2^b$ with $b\geq a$, then the number of bits
to store the integers (locations) is
$$\sum_{j=0}^{\log_2 q-1} 2^j \log_2\frac{qm}{2^j}=(q-1)ab-q(a-2)-2.$$

For instance, if $q=2^3=8$ and $m=2^7=128$, then $k=1024$ and it requires $137$ bits to represent the locations. These bits can be stored in $46$ cells without
balancing.

In fact, the above idea can be generalized for an arbitrary $q>2$. For instance, when $q=3$, given an binary word $\mathbf{u}\in G_3^{3m}$, there exists
an integer $i$ such that $\mathbf{u}+\mathbf{1}^i \mathbf{0}^{3m-i}$ has exactly $m$ $0$s or $m$ $1$s. Without loss of generality, we assume that it has exactly $m$ $0$s, then
we can further balance the subsequence consisting of $1$s and $2$s. Finally, we can get a balanced word with alphabet size $3$.
More generally, we have the following result.

\begin{Theorem}
Given an alphabet size $q=\alpha\beta$ with two integers $\alpha$ and $\beta$, we divide all the levels into $\beta$ groups, denoted by
$\{0,\beta, 2\beta, ...\}$, $\{1,\beta+1, 2\beta+1, ...\}$, ..., $\{\beta-1,2\beta-1, 3\beta-1, ...\}$.
Given any word $\mathbf{u}\in G_q^{qm}$, there exists an integer $i$ such that $\mathbf{u}+\mathbf{1}^i \mathbf{0}^{qm-i}$
has exactly $\alpha m$ symbols in one of the first $\beta-1$ groups.
\end{Theorem}

\proof Let us denote all the groups as $S_0, S_1, ..., S_{\beta-1}$. Given a sequence $\mathbf{u}$,
we use $n_j$ denote the number of symbols in $\mathbf{u}$ that belong to $S_j$.
Furthermore, we let $n_j'$ denote the number of symbols in $\mathbf{u}+\mathbf{1}^{qm}$ that belong to $S_j$.
It is easy to see that $n_{j+1}'=n_j$ for all $j\in\{0,1,...,\beta-1\}$, where $(\beta-1)+1=0$.
We prove that that there exists $j\in\{0,1,...,\beta-2\}$ such that $n_j\geq \alpha m\geq n_j'$ or $n_j\leq \alpha m\leq n_j'$ by contradiction.
Assume this statement is not true, then either $\min(n_j,n_j')>\alpha m$ or $\max(n_j,n_j')<\alpha m$ for all $j\in \{0,1,...,\beta-2\}$.
So if $n_1>\alpha m$, we can get $n_j>\alpha m$ for all $j\in \{0,1,...,\beta-1\}$ iteratively. Similarly, if $n_1<\alpha m$, we can get
$n_j<\alpha m$ for all $j\in \{0,1,...,\beta-1\}$ iteratively.  Both cases contradict with the fact that $\sum_{j=0}^\beta n_j= \alpha m \beta= qm$.

Note that the number of symbols in $\mathbf{u}+\mathbf{1}^i \mathbf{0}^{qm-i}$ that belong to $S_j$ changes by at most $1$ if we increase $i$ by one.
So if there exists $j\in\{0,1,...,\beta-2\}$ such that $n_j\geq \alpha m\geq n_j'$ or $n_j\leq \alpha m\leq n_j'$, there always exists an integer $i$ such that $\mathbf{u}+\mathbf{1}^i \mathbf{0}^{qm-i}$
has exactly $\alpha m$ symbols in $S_j$.

This completes the proof.
\hfill\QED

Based on the above result, given any $q$, we can always split all the levels into two groups and make them being balanced (the number of symbols belonging to
a group is proportional to the number of levels in that group). Then we can balance the levels in each group. Iteratively,
all the levels will be balanced.
In order to recover the original message, it requires roughly $$(q-1)\log_2 q \log_2 m$$ bits for storing additional information when $m$ is large.
If we store this additional information as a prefix using a shorter balanced code, then we get a generalized construction of Knuth's code.
If we follow the steps in Section \ref{section_bal_variant} by further adding parity-check bits, then
we get a  partial-balanced code with error-correcting capability, based on which
we can implement partial-balanced modulation for multiple-level cells.

Now, if we have  a code that uses `full' sets of balanced codewords, then the redundancy is
$$\log_2 q^{qm}-\log_2 \nchoosek{qm}{m,m,...,m}\simeq \frac{q-\log_2 q}{2}\log_2 m$$ bits.
So given an alphabet size $q$, the redundancy of the above method is about $\frac{2(q-1)\log_2 q }{q-\log_2 q}$ times as high as that of codes that uses `full' sets of balanced codewords.
For $q=2,3,4,5,...,10$, we list these factors as follows:
$$2.0000,4.4803,6.0000,6.9361,7.5694,$$
$$8.0351,8.4000,8.6995,8.9539.$$
It shows that as $q$ increases, the above method becomes less information efficient. How to construct balanced codes for a nonbinary alphabet in a simple, efficient and computationally fast way is still an open question. It is even more
difficult to construct balanced error-correcting codes for nonbinary alphabets.

\section{Conclusion}

In this paper, we introduced balanced modulation for reading/writing in nonvolatile memories.
Based on the construction of balanced
codes or balanced error-correcting codes, balanced modulation can minimize the effect of asymmetric noise, especially those
introduced by cell-level drifts. Hence, it can significantly reduce the bit error rate in nonvolatile memories.
Compared to the other schemes, balanced modulation is easy to be implemented in the current memory systems and
it does not require any assumptions about the cell-level distributions, which makes it very practical.
Furthermore, we studied the construction of balanced error-correcting codes, in particular, balanced LDPC codes.
It has very efficient encoding and decoding algorithms, and it is more efficient than prior construction of balanced error-correcting codes.

%

%
%
%





\begin{thebibliography}{1}

\bibitem{Al-Bassam90}
S. Al-Bassam and B. Bose, ``On balanced codes," \emph{IEEE Trans. Inform. Theory}, vol. 36, pp. 406--408, Mar. 1990.

\bibitem{Bez03}
R. Bez, E. Camerlenghi, A. Modelli, and A. Visconti, ``Introduction to flash memory," \emph{Proceedings of the IEEE},
vol. 91, pp. 489--502, 2003.

\bibitem{Brewer2008}
J. E. Brewer and M. Gill, \emph{Nonvolatile Memory Technologies with
Emphasis on Flash}, John Wiley \& Sons, Hoboken, New Jersey, 2008.

\bibitem{Cai2012}
Y. Cai, E. F. Haratsch, O. Mutlu, K. Mai, ``Error patterns in MLC NAND Flash memory: Measurement, characterization, and analysis," in \emph{Proc. Design, Automation, and Test in Europe (DATE)}, 2012.

\bibitem{Cover73}
T. M. Cover, ``Enumerative source coding," \emph{IEEE Trans. Inform. Theory}, vol. 19, no. 1, pp.
73--77, Jan. 1973.

\bibitem{Feldman2005}
J. Feldman, M. J. Wainwright, and D. R. Karger, ``Using linear programming to decode
binary linear codes", \emph{IEEE Trans. Inform. Theory}, vol. 51, pp. 954--972, Mar. 2005.

\bibitem{Gallager62}
R. Gallager, ``Low density parity check codes," \emph{IRE Trans. Inform. Theory},
vol. 8, no. 1, pp. 21--28, Jan. 1962.

\bibitem{Gallager63}
R. Gallager, \emph{Low Density Parity Check Codes}, no. 21 in Research Monograph
Series. Cambridge, MA: MIT Press, 1963.

\bibitem{Immink2010}
K. S. Immink and J. Weber, ``Very efficient balanced codes," \emph{IEEE Journal on Selected Areas in Communications},
vol. 28, pp. 188--192, 2010.

\bibitem{Knuth86}
D. E. Knuth, ``Efficient balanced codes," \emph{IEEE Trans. Inform. Theory},
vol. 32, no. 1, pp. 51--53, 1986.

\bibitem{Lue2008}
H. T. Lue et al., ``Study of incremental step pulse programming (ISPP)
and STI edge effect of BE-SONOS NAND flash," in \emph{Proc. IEEE Int. Symp.
on Reliability Physics}, pp. 693--694, May 2008.

\bibitem{Mazumdar2009}
A. Mazumdar, R. M. Roth, and P. O. Vontobel, ``On linear balancing
sets," in \emph{Proc. IEEE Int. Symp. Information Theory}, pp. 2699--2703, 2009.

\bibitem{McEliece98}
R. McEliece, D. MacKay, and J. Cheng, ``Turbo decoding as an instance
of Pearl's belief propagation algorithm," \emph{IEEE J. Sel. Areas Commun.},
vol. 16, no. 2, pp. 140--152, Feb. 1998.

\bibitem{Mielke2008}
N. Mielke, T. Marquart, N. Wu, J. Kessenich, H. Belgal, E. Schares, F. Trivedi, E. Goodness, and L. R. Nevill, ``Bit error rate in NAND Flash memories," in \emph{IEEE
International Reliability Physics Symposium}, pp. 9--19, 2008.

\bibitem{Parovano04}
A. Pirovano, A. Redaelli, et al., ``Reliability study of phase-change nonvolatile memories," \emph{IEEE Transactions on Device and Materials Reliability}, vol. 4, pp. 422--427, 2004.

\bibitem{Ryabko2000}
B. Y. Ryabko and E. Matchikina, ``Fast and efficient construction of an unbiased random sequence," \emph{IEEE Trans. Inform. Theory}, vol. 46, pp. 1090--1093, 2000.

\bibitem{Tallini96}
L. G. Tallini, R. M. Capocelli, and B. Bose, ``Design of some new
balanced codes," \emph{IEEE Trans. Inform. Theory}, vol. 42, pp. 790--802, May
1996.

\bibitem{vanTilborg89}
H. van Tilborg and M. Blaum, ``On error-correcting balanced codes,"
\emph{IEEE Trans. Inf. Theory}, vol. 35, no. 5, pp. 1091--1095, Sep. 1989.

\bibitem{Web2010}
J. H. Weber and K. A. S. Immink, ``Knuth's balanced code revisited,"
\emph{IEEE Trans. Inform. Theory}, vol. 56, no. 4, pp. 1673--1679, Apr. 2010.

\bibitem{Weber2012}
J. Weber, K. S. Immink and H. Ferreira, ``Error-correcting balanced Knuth codes," \emph{IEEE Trans. Inform. Theory},
vol. 58, no. 1, pp. 82--89, 2012.

\bibitem{Wong2010}
H. Wong, S. Raoux, S. Kim, J. Liang, J. P. Reifenberg, B. Rajendran, M. Asheghi, and K. E. Goodson, ``Phase change memory," \emph{Proc. IEEE},
vol. 98, no. 12, pp. 2201--2227, Dec. 2010.

\end{thebibliography}
\end{document}